\documentclass[sigconf]{acmart}

\usepackage{amsmath}
\usepackage{algorithm}
\usepackage[scientific-notation=true]{siunitx}
\usepackage{array}
\usepackage{mdwmath}
\usepackage{mdwtab}
\usepackage{eqparbox}
\usepackage{color}
\usepackage{graphicx}
\usepackage{hyperref}
\usepackage{graphicx}
\usepackage[T1]{fontenc}
\usepackage{color}
\usepackage{booktabs}
\usepackage{amsmath}
\usepackage{multirow}
\usepackage{array}
\usepackage[noend]{algpseudocode}
\usepackage{amsfonts}
\usepackage{soul}
\usepackage{subcaption}
\usepackage[compact]{titlesec}
\titlespacing\section{3pt}{3pt}{3pt}
\titlespacing\subsection{2pt}{2pt}{2pt}
\titlespacing\subsubsection{1pt}{1pt}{1pt}
\usepackage{caption}
\newcolumntype{L}[1]{>{\raggedright\let\newline\\\arraybackslash\hspace{0pt}}m{#1}}
\newcolumntype{C}[1]{>{\centering\let\newline\\\arraybackslash\hspace{0pt}}m{#1}}
\newcolumntype{R}[1]{>{\raggedleft\let\newline\\\arraybackslash\hspace{0pt}}m{#1}}

\AtBeginDocument{%
  \providecommand\BibTeX{{%
    \normalfont B\kern-0.5em{\scshape i\kern-0.25em b}\kern-0.8em\TeX}}}

\renewcommand\footnotetextcopyrightpermission[1]{}

\begin{document}

\title{Advancing the distributed Multi-GPU ChASE library through algorithm optimization and NCCL library}
\author{Xinzhe Wu}
\email{xin.wu@fz-juelich.de}
\orcid{0000-0001-5716-3116}
\affiliation{%
  \institution{J\"ulich Supercomputing Centre}
  \institution{Forschungszentrum J\"ulich GmbH}
  \streetaddress{Wilhelm-Johnen-Straße}
  \city{J\"ulich}
  \country{Germany}
  \postcode{52428}
}

\author{Edoardo Di Napoli}
\email{e.d.napoli@fz-juelich.de}
\orcid{0000-0001-5821-5897}
\affiliation{%
  \institution{J\"ulich Supercomputing Centre}
  \institution{Forschungszentrum J\"ulich GmbH}
  \streetaddress{Wilhelm-Johnen-Straße}
  \city{J\"ulich}
  \country{Germany}
  \postcode{52428}
}

\renewcommand{\shortauthors}{Wu and Di Napoli}

\begin{abstract}
As supercomputers become larger with powerful Graphics Processing Unit (GPU), traditional direct eigensolvers struggle to keep up with the hardware evolution and scale efficiently due to communication and synchronization demands. Conversely, subspace eigensolvers, like the Chebyshev Accelerated Subspace Eigensolver (ChASE), have a simpler structure and can overcome communication and synchronization bottlenecks. ChASE is a modern subspace eigensolver that uses Chebyshev polynomials to accelerate the computation of extremal eigenpairs of dense Hermitian eigenproblems. In this work we show how we have modified ChASE by rethinking its memory layout, introducing a novel parallelization scheme, switching to a more performing communication-avoiding algorithm for one of its inner modules, and substituting the MPI library by the vendor-optimized NCCL library. The resulting library can tackle dense problems with size up to $N=\mathcal{O}(10^6)$, and scales effortlessly up to the full 900 nodes---each one powered by 4$\times$A100 NVIDIA GPUs---of the JUWELS Booster hosted at the J\"ulich Supercomputing Centre.
\end{abstract}

\ccsdesc[500]{Computing methodologies~Parallel algorithms}
\ccsdesc[500]{Mathematics of computing~Mathematical software performance}

\keywords{Subspace iteration eigensolver,
Dense Hermitian matrix,
Chebyshev polynomial,
Communication-Avoiding, CholeskyQR, condition number estimation, multi-level parallelism.}


\maketitle

\section{Introduction}\label{sec:introduction}

Eigenproblems are ubiquitous in many distinct application domains of scientific computing. Algebraic eigenproblems also come in many different flavors, from dense to sparse and from symmetric to complex valued. The sought after solution can also vary widely ranging from the full eigenspectrum to just few eigenvalues in a small interval. Because of this variety, no single algorithm can solve for all the possible eigenproblem flavors. This paper describes the advances made in the ChASE library, an iterative eigensolver targeting the {\it extremal} portion of the spectrum of {\it dense Hermitian} eigenproblems ($\lambda \in [\lambda_1 , \lambda_{\sf nev}] \subset \mathbb{R}$ and $\lambda_1 < \lambda_{\sf nev} < \lambda_N$)
\[
A x = \lambda x \quad \textrm{with} \quad A^H = A \in \mathbb{C}^{N\times N}.
\]

In particular, we showcase how the recent changes in the algorithm and its parallelization scheme enables ChASE to take full advantage of some of the largest heterogeneous supercomputing platforms to solve for dense eigenproblems with size $N$ up to $\mathcal{O}(10^6)$. Some of the biggest applications where these types of problems need to be solved for the extremal portion of the spectrum is Condensed Matter Physics and Quantum Chemistry. Indeed, ChASE was originally developed to tackle problems emerging from the standard model of Condensed Matter, namely Density Functional Theory (DFT). Contrary to the standard lore which dictates that dense problems should in general be solved by direct methods (e.g., ScaLAPACK \cite{collins1969eigenvalue}, ELPA \cite{marek2014elpa,Yu2021GPU-accelerationEigenproblems}), ChASE takes the opposite path and uses an iterative method leveraging on the frequent use of the workhorse of numerical linear algebra: the {\tt GEMM} subroutine. The rational for this choice was the ability of an iterative algorithm to be inputted approximate solutions which are available in DFT computations \cite{di2012correlations}. 

In recent years, ChASE evolved to go beyond its application to eigenproblems in DFT, whose size rarely go beyond few tens of thousands, and it has been adapted to distributed heterogeneous GPU platforms to tackle problems of the size of few hundreds thousands \cite{zhang2021solving,wu2022chase}. In the course of this evolution, the library encountered a number of shortcomings which precluded its further use to more challenging cases. Due to its initial communication-avoiding design, some of the kernels (e.g., QR-factorization) which were executed redundantly on each MPI rank became the new bottleneck. In addition, the quadratic increase in the memory footprint of the internal buffers hampered the scaling beyond a couple of hundreds of nodes.

In this work we report on a number of improvements we made on ChASE. They include: i) a redesign of the buffer structure so as to move from computations executed redundantly on each process to an MPI parallelization over one of the dimensions of the 2-dimensional MPI grid; ii) a switching from a Householder QR factorization to a communication-avoiding (CA) CholeskyQR algorithm; iii) a mechanism to avoid the instabilities introduced by the CholeskyQR based on accurate estimates of the condition number of the filtered vectors; iv) a substitution of the MPI library by NCCL (NVIDIA Collective Communications Library) \cite{jeaugey2017nccl} for collective communications. In fact, the novel parallelization scheme and the employment of NCCL altogether avoid most data movement between host and device memory, which makes ChASE a library capable to easily execute at scale on the third largest accelerated cluster in Europe equipped with four NVIDIA A100 GPU cards per node.

{\it Organization.} In Section \ref{sec:background}, we give a short overview of the ChASE algorithm and the parallel implementation of the version preceding this work, followed by a description of its limitations. In Section \ref{sec:opt}, we present a newly designed parallel implementation to overcome these limitations. The numerical and parallel performance and a comparison with the currently available eigensolvers on distributed multi-GPUs architectures are illustrated in Section \ref{sec:results}. Section \ref{sec:conclusion} summarizes the achievements and concludes the paper.
\section{Background and Related Work}\label{sec:background}

ChASE (\url{https://github.com/ChASE-library/ChASE}) is a numerical library written in C++, templated for complex/real type and double/single precision and based on the subspace iteration algorithm. The subspace iteration algorithm is one of the earliest iterative methods for solving Symmetric/Hermitian eigenproblems \cite{bauer1957verfahren}. This type of algorithm with a Chebyshev polynomial filter has typically been used to solve electronic structure eigenproblems \cite{rutishauser1970simultaneous, zhou2006parallel,zhou2014chebyshev}.
Recently, ChASE library has evolved from one of these efforts to become a full-fledged numerical eigensolver that can also be used outside the electronic structure domain.


\subsection{ChASE algorithm}
 ChASE's algorithm is inspired by the work of Rutishauser \cite{rutishauser1970simultaneous} and Zhou et al. \cite{zhou2006parallel}, and features several additional components: It has an internal loop iterating over the Chebyshev polynomial filter and the Rayleigh-Ritz projection to the subspace quotient; It implements a Density of States (DoS) method to determine spectral bounds of the search subspace; It contains a deflation and locking mechanism within the internal loop. One of the most important features of ChASE is the optimization of the degree of the polynomial filter so as to minimize the number of matrix-vector operations (MatVecs) required to achieve convergence of the desired eigenpairs.
\setlength{\textfloatsep}{0pt}
\setlength{\intextsep}{3pt}
\begin{algorithm}[htbp]
    \footnotesize
  \caption{ChASE algorithm 
  }\label{alg:chase}
  \begin{algorithmic}[1]
    \Require {Hermitian matrix $H$, number of desired eigenpairs {\sf nev}, threshold tolerance for residuals {\sf tol}, initial polynomial degree {\sf deg}, search space increment {\sf nex}, vector $\hat{V}\equiv[\hat{v}_1,\cdots,\hat{v}_{{\sf nev}+{\sf nex}}]$.}
    \Ensure {{\sf nev} extreme eigenpairs $(\Lambda, \hat{Y})$}
    \State ${\sf degrees}[1:{\sf nev}+{\sf nex}]\leftarrow {\sf deg}$
    \State $(b_{sup},\mu_1,\mu_{{\sf nev}+{\sf nex}})\leftarrow$ \textsc{Lanczos}$(H)$ \label{alg:chase:lanczos}
    
    \While {size($\hat{Y}<${\sf nev})}\label{alg:chase:while}
        \State $\hat{V}\leftarrow$ \textsc{Filter}($H, b_{sup}, \mu_1, \mu_{{\sf nev}+{\sf nex}},\hat{V}, {\sf degrees}$)\label{alg:chase:filter}
        \State $\hat{Q}\leftarrow$ \textsc{QR}($[\hat{Y} \, \hat{V}]$)\label{alg:chase:qr}
        \State $(\hat{V},\tilde{\Lambda})\leftarrow$ \textsc{Rayleigh-Ritz}$(H,\hat{Q})$\label{alg:chase:rr}
        \State Compute the \textsc{Residual} $Res(\hat{V},\tilde{\Lambda})$\label{alg:chase:resid}
        \State $(\hat{V}, \Lambda, \hat{Y})\leftarrow$ \textsc{Deflation \& Locking}($\hat{V},$$\tilde{\Lambda},$ $Res(\hat{V},\tilde{\Lambda}),$$\hat{Y}$)\label{alg:chase:deflated}
        \State $\mu_1\leftarrow\min([\Lambda \, \hat{\Lambda}])$, $\mu_{{\sf nev}+{\sf nex}}\leftarrow\max([\Lambda \, \hat{\Lambda}])$
        \State $c\leftarrow\frac{b_{sup}+\mu_{{\sf nev}+{\sf nex}}}{2}$, $e\leftarrow\frac{b_{sup}-\mu_{{\sf nev}+{\sf nex}}}{2}$
        \State ${\sf degrees}[:]\leftarrow \textsc{degreeOpt}({\sf tol}, Res[:],\Lambda[:],c,e)$\label{alg:chase:degrees}
        \State \textsc{Sort} $Res(\hat{V},\tilde{\Lambda})$, $\hat{V}$, $\tilde{\Lambda}$ according to {\sf degrees}
    \EndWhile\label{alg:chase:endwhile}
  \end{algorithmic}
\end{algorithm}

Algorithm \ref{alg:chase} gives a high level description of ChASE main parts. To gain a more comprehensive understanding, we refer the reader to \cite{winkelmann2019chase}. The ChASE library first estimates the necessary spectral bounds by executing a small number of repeated Lanczos steps (Line \ref{alg:chase:lanczos}). It then filters a number of (only initially random) vectors using an optimized degree for each vector (Line \ref{alg:chase:filter}), and orthonormalizes the filtered vectors using QR factorization (Line \ref{alg:chase:qr}). The Q factor is used to reduce the original large eigenproblem to the size of the subspace through a Rayleigh-Ritz projection (Line \ref{alg:chase:rr}). The resulting "small" eigenproblem is solved using a standard dense solver such as Divide\&Conquer \cite{tisseur1999parallel}. Residuals are then computed, and eigenpairs below the tolerance threshold are deflated and locked (Line \ref{alg:chase:deflated}). Finally, a new set of filtering degrees is computed for the non-converged vectors, and the procedure is repeated (Line \ref{alg:chase:degrees}).

\subsection{Original Distributed Implementation}

The implementation of ChASE relies on a number of numerical kernels which can be separated in dense linear algebra operations to exploit optimized BLAS/LAPACK
libraries (e.g., MKL {\cite{wang2014intel}}, OpenBLAS {\cite{xianyi2012openblas}}, BLIS {\cite{van2015blis}}), and cuBLAS/cuSOLVER for GPU builds. In ChASE, MPI processes are organized as a 2D grid whose shape is as square as possible. The Hermitian matrix $H$ is distributed either following a block distribution or a block-cyclic distribution. 

ChASE most significant kernel is the Hermitian Matrix-Matrix Multiplications ({\tt HEMMs}) which, in previous versions of the library \cite{winkelmann2019chase,wu2022chase}, has been implemented for hybrid distributed-memory architectures. When NVIDIA GPUs are available, the local computation of {\tt HEMMs} on each MPI process are offloaded to the corresponding GPU(s). The {\tt HEMM} is implemented with a custom MPI scheme, and is used in the \textsc{Filter}, \textsc{Rayleigh-Ritz}, and \textsc{Residual} parts of the ChASE Algorithm. 
For instance, in the Chebyshev {\sc Filter}, which is the most computationally intensive kernel of ChASE, the matrix-matrix multiplications appears as a three-terms recurrence relation

\begin{equation}
    \label{eq:three-terms-recurrence}
    \hat{C}_{i+1} = \alpha_i(H-\gamma_i I_n)\hat{C}_{i} + \beta_i \hat{C}_{i-1}, \quad i \in [1,{\sf deg}],
\end{equation}
where $\hat{C}$ is (a subset of) the rectangular matrix $\hat{V}$, {\sf deg} is the degree of the Chebyshev polynomial, and $\alpha_i$, $\beta_i$, $\gamma_i$ are scalar parameters related to each iteration. 

A customized MPI scheme was proposed in \cite{winkelmann2019chase}, in which the rectangular matrix $\hat{C}$ is distributed over the rows of each column communicator, which also keeps a copy of $\hat{C}$. Alternatively, when $H$ is assigned to MPI processes in block-cyclic fashion, $\hat{C}$ is distributed on each column communicator using the same block size for the distribution of the rows of $H$. After a series of {\tt HEMMs} are performed in the Chebyshev {\sc Filter}, $\hat{C}$ should be re-distributed from the iteration $i$ to $i+1$. In ChASE, this re-distribution is avoided because $\hat{H}$ is symmetric/Hermitian and the $H\hat{C}$ {\tt HEMM} can be replaced by an {\tt HEMM} on $H^H\hat{C}$ when the iteration number $i$ in Equation (\ref{eq:three-terms-recurrence}) is even. This customized MPI scheme provides for an extreme good parallel performance  for {\tt HEMMs} in the {\sc Filter}.

The other operations, such as QR factorization and the eigen-decomposition within the \textsc{Rayleigh-Ritz} projection, were implemented with vendor-optimized BLAS/LAPACK by collecting redundantly a distributed matrix of vectors on each MPI process.
The QR factorization is offloaded to GPU devices through calls to the corresponding cuSOLVER functions.

\subsection{Limitations}\label{sec:limitations}

A number of concerns were reported in \cite{wu2022chase}, when ChASE v1.2 was ported to distributed multi-GPUs architectures. First, the code has large memory footprint limiting its use for very large problems. The large memory usage originates from the redundant execution on each MPI process of some of the numerical kernels, including \textsc{QR}, the \textsc{Rayleigh-Ritz} projection and the computation of \textsc{Residual}s.
The redundant computation requires the storing of two buffers of size $\mathcal{O}{(N({\sf nev}+{\sf nex}))}$ on the physical memory assigned to each MPI process. 
Because the memory device is limited compared to the main memory, this limitation is particularly severe in the distributed multi-GPUs build of ChASE. 
Similarly, the redundant execution of a portion of the computation also results in a non-scalable part (mainly linked to the QR factorization) of the distributed ChASE.
This limitation has a direct influence on the maximum size of matrix $H$ that could be tackled and the number of eigenpairs that could be computed. 

A second concern regards the communication overhead for the kernels executed redundantly on each MPI process. These kernels need the collection of a distributed matrix of vectors within the row (or column) communicator into a redundant buffer on each task. The collection is obtained by the individual broadcasting of a buffer for each task within the row (or column) communicator. When the count of MPI tasks quadruples, the number of messages doubles. As shown in Fig. \ref{fig:Computation vs Communication}, this limitation particularly harms the weak scaling performance of ChASE, when the word size per message is fixed while the the number of messages increases. 

When using accelerators, version v1.2 follows a host-device mode which offloads the most computationally intensive operations to the GPU devices. Once the computation is completed, the related data are immediately copied back to CPUs introducing a substantial data movement overhead between host and device memory. The use of traditional MPI libraries for the collective communications requires further data movement from GPU to CPU once communication operations are invoked.   

In summary, ChASE v1.2 suffers from the layout of its parallelization scheme, in which some numerical kernels are redundantly executed on each MPI process. These limitations were insignificant when ChASE was initially designed because it was mainly targeting problems from electronic structure calculations, whose size typically ranges from a few thousands to several tens of thousands. However, when the ambition of the ChASE library grew to encompass much larger problems, its parallel structure became the source of several bottlenecks. Porting ChASE to GPUs, which are extremely powerful for the computation of standard linear algebra operations, amplified further these shortcomings.   
\section{Optimization and Implementation}\label{sec:opt}
In this paper, we present a novel scheme for ChASE that parallelizes \textsc{QR}, \textsc{Rayleigh-Ritz} and \textsc{Residual} on a subset of the MPI grid---either row or column communicators---and in doing so removes the need of large and redundant buffers on each MPI process. Importantly, it enables a distributed multi-GPU implementation to avoid almost all host-device data movement by keeping the computations of all major numerical kernels on the GPUs and communicating data via NCCL. Moreover, we introduce new faster QR algorithms and a mechanism to select the most appropriate based on the condition number of the filtered vectors.  

\subsection{Novel Parallelization Scheme}
In this section, we present the novel parallelization scheme in a general way which is applicable for both CPU and GPU builds. As shown in Algorithm \ref{alg:new}, a Hermitian matrix $H$ is distributed onto a 2D MPI grid following either a block distribution or a block-cyclic distribution, and the local block on each MPI process is of size $n_r \times n_c$. Arrays $C, C_2$ and $B, B_2$ of size, respectively $n_r\times n_e$ and $n_c\times n_e$, are allocated on each MPI process, with $n_e={\tt nev} + {\tt nex}$. The set of all arrays $C$ and $C_2$ represent a rectangular matrix of size $N \times n_e$ within each column communicator, while the set of all $B$ and $B_2$ represent a matrix of same size within each row communicator. The entries of the global matrices represented by $C$ and $C_2$ are identical among different column communicators, and $B$ and $B_2$ are identical among row communicators. $C$ is designed as the buffer which receives the initial input vectors 
and returns the computed eigenvectors after completing the execution.

The implementation of \textsc{Filter} remains unchanged. If the iteration index in Equation (\ref{eq:three-terms-recurrence}) is odd, the algorithm performs $HC$ and stores the result in $B$. If the index is even, the algorithm executes $H^HB$ and writes the result to $C$. As ChASE enforces even-degree Chebyshev polynomials, the filtered vectors are always stored in $C$. At the next step, a 1D parallelized QR factorization is performed on $C$ within each column communicator. We introduce a mechanism that can switch between different communication-avoiding QR (CAQR) implementations to strike a balance between optimal performance and good numerical stability. Details of the CAQR method are provided in Section \ref{sec:caqr}. After the QR factorization, the first {\sf locked} (converged) columns in $C$ are replaced with the corresponding columns stored in $C_2$ at the previous iteration, while the remaining columns of $C_2$ are replaced with the corresponding newly orthogonalized columns of $C$.

\begin{algorithm}
    \footnotesize
	\caption{ChASE with Distributed QR factorization} \label{alg:new}
	\begin{algorithmic}[1]
	    \Require 
            {\sf nev}: number of eigenpairs to be computed; 
            {\sf nex}: search space increment;
	    {\sf $n_e$ = nev + nex}; 
            \textsf{rcomm} \& \textsf{ccomm}: row and column communicator of 2D MPI grid;
	    $H$: {$n_r$}$\times${$n_c$} dense matrix; 
            $C$ \& $C_2$: {$n_r$}$\times${\sf $n_e$} dense matrix; 
            $B$ \& $B_2$: {$n_c$}$\times${\sf $n_e$} dense matrix; 
            $A$: {\sf $n_e$}$\times${\sf $n_e$} dense matrix; 
            {\sf MaxIter}: maximal iteration number; 
            {\sf deg}: degree of filter; 
            {\sf tol}: threshold tolerance for residuals;
            {\sf opt}: indicating if degree optimization enabled
	    \Ensure first {\sf nev} lowest eigenpairs of matrix represented by {\sf H} within 2D MPI grid.
            \State $(b_{sup},\mu_1,\mu_{n_e})\leftarrow$ \textsc{Lanczos}$(H)$
	    \State {\sf iter} = 1, {\sf locked} = $0$, {\sf degs}[:]$=${\sf deg}, $\Lambda[:]=0$
            \State $c\leftarrow\frac{b_{sup}+\mu_{n_e}}{2}$, $e\leftarrow\frac{b_{sup}-\mu_{n_e}}{2}$
	    \While {{\sf locked } $<$ {\sf nev} \&\& {\sf iter} $\leq$ {\sf MaxIter}  }
	        \If {{\sf iter} $\neq 1$  }
                    \State $\mu_1, \mu_{n_e} \leftarrow$\textsc{updateBounds}({$\Lambda$})
                    \State $c\leftarrow\frac{b_{sup}+\mu_{n_e}}{2}$, $e\leftarrow\frac{b_{sup}-\mu_{n_e}}{2}$
                    \If{{\sf opt}}
                    \State ${\sf degs}[:]\leftarrow \textsc{degreeOpt}({\sf tol}, {\sf Res}[:], \Lambda[:], c, e)$
                    \EndIf
	        \EndIf
	        \State $C[:,{\sf locked}+1:]\leftarrow$\textsc{Filter}($H$, $C[:,{\sf locked}+1:]$, {\sf degs}, $c$) \label{Alg:new:line:filter}
                \State {\sf cond}$\leftarrow${\sc CondEst}($\Lambda$, $c$, $e$, {\sf degs}, {\sf locked})
                \State $C\leftarrow$\textsc{1D-CAQR}($C$, {\sf cond}, {\sf ccomm}) \label{Alg:new:line:caqr}
	        \State $C[:, 1:{\sf locked}]\leftarrow C_2[:, 1:{\sf locked}]$, $C_2[:, {\sf locked}+1:]\leftarrow C[:, {\sf locked}+1:]$\label{Alg:new:line:copy1}
	        \State $B2\leftarrow\textsc{Bcast}(C_2, {\sf ccomm})$
                \State $B[:, {\sf locked}+1:]\leftarrow HC[:, {\sf locked}+1:]$\label{Alg:new:line:H*C}
                \State $A\leftarrow (B_2[:,  {\sf locked}+1:])'B[:, {\sf locked}+1:]$\label{Alg:new:line:B*B2}
	        \State $A\leftarrow$\textsc{AllReduce}$(A, \textsc{SUM}, \textsf{rcomm})$
	        \State $\Lambda, A\leftarrow\textsc{HE(SY)EVD}(A)$ \label{Alg:new:line:heevd}
	        \State $C[:, {\sf locked}+1:]\leftarrow C_2[:, {\sf locked}+1:]A$\label{Alg:new:line:C*A}
                \State $C_2[:, {\sf locked}+1:]\leftarrow C[:, {\sf locked}+1:]$ \label{Alg:new:line:copy3}
	        \State $B2\leftarrow\textsc{Bcast}(C_2, {\sf ccomm})$
                \State $B[:, {\sf locked}+1:]\leftarrow HC[:, {\sf locked}+1:]$\label{Alg:new:line:H*C2}
                \State $B[:, {\sf locked}+1:]\leftarrow B[:, {\sf locked}+1:] -{\sf ritzv}[{\sf locked}+1:]B_2[:,{\sf locked}+1:]$\label{alg:new:line:axpy}
                \State \textsf{nrm}[{\sf locked}+1:]$\leftarrow$\textsc{SquaredNorm}($B[:, {\sf locked}+1:]$)\label{alg:new:line:sqrtnorm}
	        \State \textsf{nrm}$\leftarrow$\textsc{AllReduce}$(\textsf{nrm}, \textsc{SUM}, \textsf{rcomm})$
                \State \textsf{resd}$[{\sf locked}+1:]\leftarrow \sqrt{\textsf{nrm}[{\sf locked}+1:]}$ \label{alg:new:line:resd}	        
	        \State $C$, $C_2$, {\sf new\_converged}$\leftarrow$\textsc{Locking}($C$, $C_2$, \textsf{resd}, {\sf tol})
	        \State {\sf locked}=${\sf locked}+${\sf new\_converged}, {\sf iter}=${\sf iter}+1$
         
            \EndWhile
            \State \Return: $\Lambda[1:{\sf nev}]$, $C[:, 1:{\sf nev}]$
	\end{algorithmic} 
\end{algorithm}

The {\sc Rayleigh-Ritz} step projects the original problem onto a {\it search} subspace, from which approximate solutions are computed. The active subspace is obtained by forming a $n_e \times n_e$ {\sc Rayleigh-Ritz} quotient $A={C}^HH{C}$, with ${C}$ the $N\times n_e$ orthonormal matrix outputted by the QR factorization, which is distributed within each column communicator. The right-multiplication of ${C}$ with $H$ is implemented by employing the same distributed {\tt HEMM} used in the {\sc Filter}. The result of $HC$ is stored in $B$, which is distributed over the row communicator. The left-multiplication of ${C}^H$ with $H{C}$, realized as $C^HB$, requires first to copy $C$ buffers from the column communicator into the $B_2$ buffers distributed within the row communicator. If the 2D MPI grid is square, then a single broadcasting operation is sufficient. 
However, if the MPI grid is non-square, multiple broadcasting operations may be required depending on the shape of the MPI grid and the way in which $H$ is distributed, especially for block-cyclic distributions. Squared MPI grids are the optimal configuration for ChASE, as with other state-of-the-art eigensolvers such as ELPA.

$C^HB$ is deployed as $B_2^HB$ in Algorithm \ref{alg:new} Line \ref{Alg:new:line:B*B2}, which is naturally parallel within the row communicator. Each $B_2^HB$ is a just a call to {\tt GEMM} on the local blocks $B$ and $B_2$ stored on each MPI process, and the result is written in the redundant buffer $A$, through an {\tt AllReduce} operation with addition operation along the row communicator. 
$A$ is then diagonalized redundantly on each MPI process as $(Y,\Lambda) \leftarrow A$ using a LAPACK eigensolver, where $\Lambda$ are the approximate eigenvalues of $A$, and $A$ is overwritten by the related eigenvectors $Y$. The eigenvectors of the original problem, which are designed to be stored in $C$, can be obtained through the back-transform $C_2A$ in Algorithm \ref{alg:new} Line \ref{Alg:new:line:C*A}, which is naturally parallel within the column communicator, since $C_2$, which is identical to $C$ in this step, is distributed, and $A$ is redundant on all processes.

The {\sc Residual} step computes the Euclidean norm of each column  of $C\Lambda-HC$, which is equivalent to computing the Euclidean norm of each column of $B_2 \Lambda-B$, as $B_2$ is re-distributed from $C_2$. The complexity of the required memory on each MPI process is

\begin{equation}
\label{eq:CPU_mem}
\begin{aligned}
M_{cpu}=\frac{N^2}{pq}+2\frac{Nn_e}{p} + 2\frac{Nn_e}{q} + {n_e^2} ,
\end{aligned}
\end{equation}
where $p\times q$ is the dimension of the 2D MPI grid.
 The first term corresponds to the local block of $H$ held by each MPI process, the second term corresponds to the $C$ and $C_2$ buffers, and the third term corresponds to the $B$ and $B_2$ buffers. The fourth and smallest term comes from the $A$ blocks.

\subsection{Communication-avoiding QR factorization}\label{sec:caqr}
In order to gain better performance, we replace the Householder QR (HHQR) by variants of communication-avoiding (CA) CholeskyQR \cite{fukaya2014choleskyqr2,fukaya2020shifted}. For each iteration in ChASE, a QR factorization is performed on a rectangular matrix, here referred to as $X$, whose size is $m\times n$ with $m>n$. We preferred CholeskyQR rather than tall-skinny QR (TSQR) \cite{demmel2008communication}, a CAQR with equivalent communication costs as CholeskyQR, because
the reduction operator of CholeskyQR is addition, while that of TSQR is the QR factorization of a small matrix \cite{fukaya2014choleskyqr2}. To our knowledge, TSQR is advantageous over ScaLAPACK-HHQR  in 1D MPI grid only if $m\gg n$ \cite{ballard2015reconstructing}. This is not the case for ChASE, which, in typical Condensed Matter problems, is expected to solve for around $10\%$ of the extremal eigenpairs.

\begin{algorithm}[b]
    \footnotesize
	\caption{1D-CholeskyQR for $X=QR$} \label{alg:cholqr}
	\begin{algorithmic}[1]
            \Function{CholeskyQR}{$X$, {\sf comm}, {\sf cholDegree}}
              \For{$i = 1, \cdots,${\sf cholDegree}}
                \State $R\leftarrow \textsc{SYRK}(X)$
                \State $R\leftarrow$\textsc{AllReduce}$(R, \textsc{SUM}, \textsf{comm})$
                \State [$R, {\sf info}$]$ \leftarrow \textsc{POTRF}$($R$)
	          \State $X \leftarrow \textsc{TRSM}$($X$, $R$)
              \EndFor
              \State \Return $X$
            \EndFunction
	\end{algorithmic} 
\end{algorithm}

Despite the better performance, CholeskyQR is afflicted by a rapid decline in orthogonality across the columns of the $Q$ factor for increasing condition number of $X$. This instability can be significantly mitigated by performing the algorithm twice, which is known as CholeskyQR2 \cite{fukaya2014choleskyqr2}. The applicability of CholeskyQR2 is still limited by the requirement that the Cholesky factorization of the Gram matrix $R = X^HX$ runs to completion, which requires the matrix $X$ to have a condition number $\kappa_2(X)$ no larger than $\mathcal{O}(\mathbf{u}^{-1/2})$, where $\mathbf{u}$ is the unit round-off and $\kappa_2(X) = \frac{\sigma_{max}(X)}{\sigma_{min}(X)}$, with $\sigma_{max}(X)$ and $\sigma_{min}(X)$ respectively its maximal and minimal singular values. This limitation has been addressed in \cite{fukaya2020shifted}, where a preconditioning step is added to CholeskyQR2 to reduce the condition number of $X$ to a point where CholeskyQR2 is applicable. This improved version is called shifted CholeskyQR2 or $s$-CholeskyQR2, and it can handle matrices with condition numbers up to $\mathcal{O}(\mathbf{u}^{-1})$.

Since computing the exact condition number of a rectangular matrix is computationally expensive, we introduce an accurate and cost-free mechanism to estimate $\kappa_2(X)$ which ultimately guides us in choosing the best QR-factorization variant for the array of filtered vectors.
This estimation is the result of a numerical analysis of the spectral properties of $X$ and 
it will appear in an upcoming manuscript. The estimate is implemented in Algorithm \ref{alg:cond_est}, using input arguments that are already available in ChASE.
In Section \ref{sec:Estimating the condition number}, we illustrate the effectiveness of our estimation methodology, while in Section \ref{sec:cholqr_vs_hhqr}, we compare CholeskyQR, with the proposed heuristic, against HHQR and show that both achieve a similar convergence behavior while CholeskyQR returns a much better performance.

A distributed-memory implementation of CholeskyQR is given in Algorithm \ref{alg:cholqr}, where $X$ represents the matrix to be factorized, and \textsf{comm} is the MPI communicator. The variable \textsf{cholDegree} specifies the number of repetitions of CholeskyQR, so, if it equals to $2$, it is indeed CholeskyQR2. For each repetition, it starts by partially calculating the Gram matrix $X^HX$ on each MPI process, through LAPACK \textsc{ZHERK} (\textsc{DSYRK}) routines. The final Gram matrix is obtained via an {\tt AllReduce} operation with addition. Then a Cholesky factorization is performed on $R$ with LAPACK \textsc{POTRF}, and $Q$ is computed via back substitution with the LAPACK \textsc{TRSM}. 
\setlength{\floatsep}{3pt}

\begin{algorithm}[t]
    \footnotesize
	\caption{Distributed 1D-CAQR for ChASE}\label{alg:caqr}
	\begin{algorithmic}[1]
            \Function{1D-CAQR}{$X$, \textsf{estCond}, \textsf{comm}} 
            \If{\textsf{estCond}$>10^8$}
                \State $R\leftarrow \textsc{SYRK}(X)$
                \State $R\leftarrow$\textsc{AllReduce}$(R, \textsc{SUM}, \textsf{comm})$
                \State ${\sf norm}\leftarrow$\textsc{AllReduce}$(||X||_F^2, \textsc{SUM}, \textsf{comm})$
                \State $s = 11(mn+n(n+1))\mathbf{u}{\sf norm}$
                \State [$R, {\sf info}$] $\leftarrow \textsc{POTRF}$($R+sI$)
                \If{${\sf info}\neq 0$}
                    \State $X \leftarrow$\textsc{ScaLAPACK-HHQR}($X$,\textsf{comm})\label{alg:line:scalapack_qr}
                \Else
                    \State $X \leftarrow \textsc{TRSM}$($X$,$R$)
                    \State $X\leftarrow \textsc{CholeskyQR}$($X$, \textsf{comm},$2$)
                \EndIf
            \ElsIf{\textsf{estCond}$<20$}
                \State $X\leftarrow \textsc{CholeskyQR}$($X$, \textsf{comm}, $1$)
            \Else
                \State $X\leftarrow \textsc{CholeskyQR}$($X$, \textsf{comm}, $2$)
            \EndIf
            \State \Return $X$            
            \EndFunction
	\end{algorithmic} 
\end{algorithm} 

\begin{algorithm}[h!]
    \footnotesize
	\caption{Condition number estimation of filtered vectors} \label{alg:cond_est}
	\begin{algorithmic}[1]
            \Function{CondEst}{$\Lambda$, $c$, $e$, {\sf degs}, {\sf locked}} 
                \State $t'=\frac{{\Lambda}[1]-c}{e}$, $t=\frac{{\Lambda}[{\sf locked}+1]-c}{e}$              \State $|\rho|=\max$\{$|t-\sqrt{t^2-1}|$, $|t+\sqrt{t^2-1}|$\}
                \State $|\rho'|=\max$\{$|t'-\sqrt{t'^2-1}|$, $|t'+\sqrt{t'^2-1}|$\}
                \State ${\sf d} = {\sf degs}[{\sf locked}+1]$,  ${\sf d}_{M} = \max({\sf degs}[{\sf locked}+1:])$
                \State ${\sf cond} = |\rho|^{{\sf d}}|\rho'|^{({\sf d}_{M}-{\sf d})}$
            \State \Return {\sf cond}
            \EndFunction
	\end{algorithmic} 
\end{algorithm}

Based on the estimated condition number of $X$, a heuristic for selecting the appropriate CholeskyQR variant is proposed as Algorithm \ref{alg:caqr}. If the estimated condition number is larger than $\mathcal{O}(\mathbf{u}^{-1/2})$, which is approximately $10^8$ for double precision, we select the $s$-CholeskyQR2 variant. If the estimated condition number is smaller than a fixed threshold (in practice set to $20$), CholeskyQR is sufficient. Otherwise, CholeskyQR2 should be used. To ensure robustness, we 
revert to HHQR in Algorithm \ref{alg:caqr} (Line \ref{alg:line:scalapack_qr}) to prevent failures of $s$-CholeskyQR2 for any corner case. 

\subsection{Porting to GPUs}

A straightforward way to port ChASE to distributed GPU clusters is to offload its most computation-intensive operations onto GPUs.  Specifically, the local computation of matrix-matrix multiplication in Line \ref{Alg:new:line:filter}, \ref{Alg:new:line:H*C}, \ref{Alg:new:line:H*C2}, \ref{Alg:new:line:C*A}, \ref{Alg:new:line:B*B2}, the \textsc{SYRK}, \textsc{POTRF} and \textsc{TRSM} of Line \ref{Alg:new:line:caqr}, the \textsc{HE(SY)EVD} in Line \ref{Alg:new:line:heevd} of Algorithm \ref{alg:new} have been ported to GPUs by using the corresponding routines provided by cuBLAS and cuSOLVER. The BLAS-1 operations of Line \ref{alg:new:line:axpy}, \ref{alg:new:line:sqrtnorm} and \ref{alg:new:line:resd} for \textsc{Residuals} stay on CPUs. The computed results of the operations on GPUs are copied back to CPUs once the computations are done.

In fact, the memory copying operations for the collective operations can be bypassed by exploring the GPUDirect technology. Specifically, we can use the optimized GPU-driven NCCL library to replace the MPI library for all the collective communications (all-reduce and broadcast) in ChASE. NCCL provides communication primitives for the collective communications. Starting from NCCL2, it has support for InfiniBand based communication and can span multiple nodes. Since the NCCL APIs are not MPI-compliant, a 2D NCCL communicator has been built on top of the 2D MPI grid in ChASE so that each MPI process is mapped to a single GPU device within this 2D NCCL communicator. If the buffers $C$, $C_2$, $B$, $B_2$ and $A$ reside on the device, {\sf ccomm} and {\sf rcomm} are the corresponding column/row communicator within the 2D NCCL communicator, and all the operations of \textsc{AllReduce} and \textsc{Bcast} are substituted by their equivalents in NCCL. This means that all the computations are executed on the GPUs by using the corresponding routines provided by cuBLAS and cuSOLVER, and all the host-device data movement for all major kernels have been eliminated. Furthermore, the Line \ref{alg:new:line:sqrtnorm} in Algorithm \ref{alg:new} for \textsc{Residuals} has been also offloaded to GPUs as a single batched kernel for a series of BLAS-1 operations. 

In this paper, we refer respectively to the implementation of ChASE with and without NCCL support as ChASE(NCCL) and ChASE(STD), where STD, referring to \textit{standard}, implies the standard way of distributed GPU communication with explicit host-device data movement.  We prefer NCCL over other \textit{CUDA-Aware} MPI libraries, such as OpenMPI and MVAPICH2, as the latter are rather designed with many optimized GPU-based point-to-point communication schemes while the former is targeting collective communications.

\section{Numerical Experiments}\label{sec:results}

ChASE has been tested on the supercomputer JUWELS-Booster at J\"ulich Supercomputing Centre in Germany, which consists of 936 NVIDIA GPU-accelerated compute nodes. The configuration of each node is two 24 cores AMD EPYC 7402 CPUs @ 2.25 GHz ($16 \times 32$ GB DDR4 Memory), 4$\times$NVIDIA A100 GPU with 40 GB memory. The interconnect are 4$\times$ InfiniBand HDR (Connect-X6).

The C/C++ compiler used is GCC 11.3.0, MPI library is OpenMPI 4.1.4, and BLAS/LAPACK libraries are Intel MKL 2022.1.0, CUDA version is 11.7. 
For ChASE(STD) and ChASE(NCCL), the number of MPI ranks per node is $4$, with $1$ GPUs and $12$ OpenMP threads per rank. ChASE v1.2 is marked as ChASE(LMS), in which LMS refers to \textit{Limited Memory and Scaling}. It is configured with $1$ MPI rank per node, with $4$ GPU and $12$ OpenMP threads per rank. The threshold tolerance for residuals {\tt tol} is fixed as
$10^{-10}$, and the degree optimization of the ChASE filter is always enabled unless otherwise specified. All tests in this paper are performed in double-precision.

\subsection{Test Matrix Suite}
For the tests we use an heterogeneous collection of eigenproblems either coming from domain applications or artificially generated.
\subsubsection{DFT and BSE matrices}\ Eigenproblems from applications are extracted from DFT and Bether-Salpeter simulations. Details of these problems are listed in Table \ref{tab:real_world matrix}, including an acronym, the size, the number of eigenpairs sought after {\sf nev}, the size of extra searching space {\sf nex}, the application software used to extract them, and the type of each problem. The \text{FLEUR} problems are generated by FLEUR \cite{fleurwebsite} code, 
The BSE UIUC problems are obtained through a fork of the Jena BSE code developed and maintained at the University of Illinois Urbana-Champaign \cite{zhang2021solving}.

\begin{table}[htbp]
\scriptsize
	\renewcommand{\arraystretch}{.9}
	\caption{List of DFT matrices for numerical tests.}\label{tab:real_world matrix}
	\centering
	\begin{tabular}{C{2cm}|cccC{1.5cm} c}
		\toprule
  Name & $N$ & {\sf nev} & {\sf nex} & Source & Type \\
  \midrule
  \textsc{NaCl 9k}   & 9273 & 256 & 60 & \textbf{FLEUR} & Hermitian \\
  \hline
  \textsc{AuAg 13k}   & 13379 & 972 & 100 & \textbf{FLEUR} & Hermitian\\
  \hline  
  \textsc{TiO2 29k}   & 29528 & 2560 & 400 & \textbf{FLEUR} & Hermitian\\
  \hline  
  \textsc{In2O3 76k}  & 76887 & 100 & 40 & \textbf{BSE UIUC}& Hermitian\\
  \hline
  \textsc{In2O3 115k}  & 115459 & 100 & 40 & \textbf{BSE UIUC}& Hermitian \\
  \hline  
  \textsc{HfO2 76k}    & 76674 & 100 & 40 & \textbf{BSE UIUC} & Hermitian\\
		\bottomrule
	\end{tabular}
\end{table}

\subsubsection{Artificial Matrices}\
For benchmarking the parallel performance of ChASE, artificial matrices are generated with a given spectrum, which is inspired by the testing infrastructure in LAPACK \cite{marques2008algorithm}. To generate them, we construct a diagonal matrix $D$ filled with the prescribed eigenvalues. Then a dense matrix $A$ with the given spectra is generated as $A=Q^TDQ$, with $Q$ the first factor of the QR factorization of a random square matrix. In this paper, the eigenvalues of the artificial matrices are distributed uniformly within an interval and will be referred to as {\sf Uniform} matrices.

\subsection{Estimating the condition number}\label{sec:Estimating the condition number}

\begin{figure}[t]
    \centering
    \begin{subfigure}{0.23\textwidth}
        \centering
        \includegraphics[width=\textwidth]{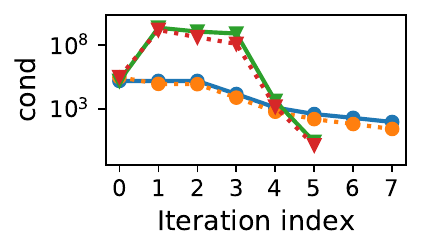} 
        \caption{NaCl 9k}
    \end{subfigure}
    \begin{subfigure}{0.23\textwidth}
        \centering
        \includegraphics[width=\textwidth]{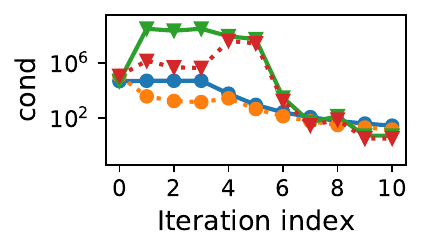}
        \caption{AuAg 13k}
    \end{subfigure}
    ~\\
    \begin{subfigure}{0.23\textwidth}
        \centering
        \includegraphics[width=\textwidth]{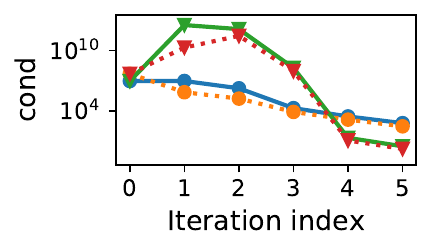} 
        \caption{TiO2 29k}
    \end{subfigure}
    \begin{subfigure}{0.23\textwidth}
        \centering
        \includegraphics[width=\textwidth]{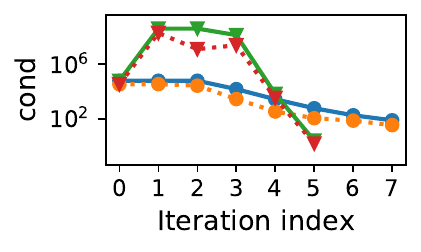}
        \caption{In2O3 76k} 
    \end{subfigure} 
    ~\\
    \begin{subfigure}{0.23\textwidth}
        \centering
        \includegraphics[width=\textwidth]{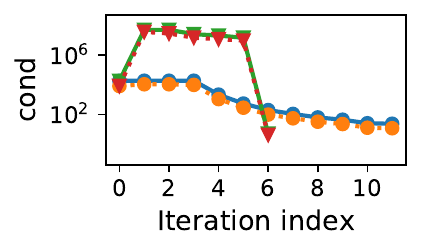} 
        \caption{In2O3 115k}
    \end{subfigure}
    \begin{subfigure}{0.23\textwidth}
        \centering
        \includegraphics[width=\textwidth]{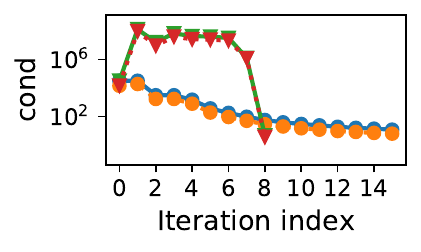}
        \caption{HfO2 76k} 
    \end{subfigure} 
    
    \caption{Comparison of $\kappa_2^{est}$ (solid line) and the accurate $\kappa_2^{com}$ condition number for each iteration of ChASE solving eigenproblems listed in Table \ref{tab:real_world matrix} with (triangular marker) and without (disc marker) degree optimization.}\label{fig:cond_est}
\end{figure}

In this section we illustrate the effectiveness of the upper bounds introduced in Section \ref{sec:caqr} for the condition number of the rectangular matrix of vectors $C$ outputted by the Chebyshev filter\footnote{We indicate with $C$ both the array stored in each MPI process and the union of such arrays representing the full matrix of vectors. The difference in usage can be easily evinced from the context.}. The test problems are the ones listed in Table \ref{tab:real_world matrix}. The results are shown in Fig. \ref{fig:cond_est} which compares the estimated condition number $\kappa_2^{est}$
against an accurate computation of the condition number $\kappa_2^{com}$
of the filtered matrices for each iterations of ChASE up to completion. We carried on the tests with the polynomial degree optimization turned either on ({\it opt}) or off ({\it no-opt}) to show how the condition number estimation intrinsically depends on the optimization mechanism. 
For the case {\it no-opt}, the Chebyshev polynomial degree is fixed to $20$ at every iteration. For the case {\it opt}, the initial degree for the first iteration is set as $20$ and changes at every later iteration for each of the filtered vectors in $C$. A maximal allowed degree is fixed to $36$ to avoid the matrix of vectors becoming too ill-conditioned.
The $\ell_2$-norm condition number $\kappa_2^{com}(C)$ 
 are computed by LAPACK SVD solvers after collecting the distributed blocks along the column communicator into a redundant matrix.

Fig. \ref{fig:cond_est} shows that $\kappa_2^{est}$, for both {\it no-opt} and {\it opt}, always bounds from above $\kappa_2^{com}$, making it an effective and reliable estimate. Exceptions may happen for the first iteration index where the estimated condition number is slightly lower than the computed one differing only in the last computed digit. This mismatch has its origin from the implicit assumption, used to derive estimation formula in Algorithm \ref{alg:cond_est}, that the condition number of the input matrix of the filter is always $1$, which is not necessarily true for arrays of randomly generated orthonormal vectors.

Often, the ratio between $\kappa_2^{est}$ and $\kappa_2^{com}$ is below $2$. For some cases (i.e., {\it opt} case for AuAg 13k) this ratio can reach at most $\mathcal{O}(10^4)$ for a few initial iterations. This is likely caused by the inaccuracy of the estimates for the parameters $e$ and $c$ during the first 2-3 iterations, where no eigenpairs have been locked yet. Independently from its strictness, $\kappa_2^{est}$ always bound $\kappa_2^{com}$ from above and so it constitutes a reliable parameter for ChASE to switch from a more stable QR factorization, such as $s$-Cholesky QR factorization, to the less stable but more efficient CholeskyQR2, even CholeskyQR1 in the last one or two iterations where the estimated condition number is $\mathcal{O}(1)$. 

In the {\it no-opt} case, the highest condition number comes at the first iteration. Therefore, if the condition number of $C$ at the first iteration is below a certain threshold, the $s$-CholeskyQR2 can be avoided in any of the following iterations. Conversely, in the {\it opt} case this condition number at the early stage can be much larger than the one at the first iteration. This effect is caused by the higher maximal degree allowed during the degree optimization procedure and it is controllable by the user. Conversely, ChASE with degree optimization can always converge much faster than ChASE without degree optimization. In conclusion, our upper bound estimation of the condition number for the filtered matrix ensure that the proposed heuristic to switch between QR variants is reliable throughout the entire execution cycle of ChASE. 

\subsection{ChASE with CholeskyQR vs with HHQR}\label{sec:cholqr_vs_hhqr}

We compare the numerical behaviour of ChASE (NCCL) equipped with HHQR (for all ChASE iterations) with the automatic selection mechanism introduced in Section \ref{sec:caqr}. Here, HHQR specifically refers to the Householder QR implementation provided by ScaLAPACK, which uses a 1D MPI grid and is executed independently over each column communicator. The block size of ScaLAPACK block-cyclic distribution for the rows is the same as the number of rows of $C$, and the block size for the columns is fixed at $32$.  Data of each test are obtained as the averages of $5$ repetitions.

The test problems are the same listed in Table \ref{tab:real_world matrix}. ChASE are tested using $4$ compute nodes on JUWELS-Booster with results reported in Table \ref{table:hh_vs_chol_gpu}. This table shows the total number of MatVec operations, the number of iterations to convergence, the total time-to-solution and the execution time of the QR factorization.

\begin{table}[t]
\scriptsize
\renewcommand{\arraystretch}{1.}
\caption{Comparison of ChASE(NCCL) with HHQR and CholeskyQR solving for eigenproblems in Table \ref{tab:real_world matrix}. The tests are performed on JUWELS-Booster with $4$ nodes.}\label{table:hh_vs_chol_gpu}
\centering
\begin{tabular}{C{1.5cm}|C{1.4cm}|C{1.2cm}C{0.5cm}C{1.1cm}C{.8cm}}
\toprule
Type                     & QR Impl.   & MatVecs & Iters & All (s) & QR (s) \\ \midrule
\multirow{2}{*}{\textsc{NaCl 9k}}
                         & HHQR   & 31,146 & 6 & 1.49 & 1.05 \\ 
                         & CholeskyQR & 31,146 & 6 & 0.43 & 0.03  \\ \cline{1-6} 
\multirow{2}{*}{\textsc{AuAg 13k}}
                         & HHQR   & 124,852 & 11 & 24.68 & 22.71 \\ 
                         & CholeskyQR & 124,852 & 11 & 10.92 & 0.20  \\ \cline{1-6} 
 \multirow{2}{*}{\textsc{TiO2 29k}}
                         & HHQR   & 213,790 & 5 & 167.39 & 157.02 \\ 
                         & CholeskyQR & 213,790 & 5 & 8.80 & 0.48 \\  \cline{1-6} 
\multirow{2}{*}{\textsc{In2O3 76k}}
                         & HHQR   & 14,818 & 6 & 9.81 & 2.26 \\  
                         & CholeskyQR & 14,818 & 6 & 7.64 & 0.13 \\  \cline{1-6} 
\multirow{2}{*}{\textsc{In2O3 115k}}
                         & HHQR   & 18,678 & 7 & 23.83 & 3.92 \\ 
                         & CholeskyQR & 186,78 & 7 & 20.16 & 0.22 \\    \cline{1-6} 
\multirow{2}{*}{\textsc{HfO2 76k}}
                         & HHQR   & 25,664 & 9 & 14.11 & 3.38 \\ 
                         & CholeskyQR & 25,664 & 9 & 10.92 & 0.20 \\   
\bottomrule
\end{tabular}
\label{tab:chol_vs_hh_gpu}
\end{table}

For all the tests, the usage of either HHQR or CholeskyQR results in the same convergence behaviour with the same number of MatVec operations and iterations. The speedup of ChASE with CholeskyQR over ChASE with HHQR is clearly noticeable. 
Moreover, the employment of CholeskyQR greatly enhances the performance for ChASE-GPU when more than $1,000$ eigenpairs are sought after.

\subsection{Kernel Profiling}\label{sec:Communication vs Computation}

In this section, we compare the communication, computation and data movement for the main parts of different implementations of ChASE: the {\sc Filter}, {\sc QR}, \textsc{Rayleigh-Ritz} and \textsc{Residuals}. We designed a weak-scaling experiment, in which the count of compute nodes increases from $1$ to $64$, while the matrix size increases from $30$k to $240$k. For this experiment, we used artificial matrices of type {\sf Uniform} with {\sf nev} and {\sf nex} being fixed to $2,250$ and $750$. Only the first iteration is reported, which ensures a fixed workload per task with the increase of compute nodes count. 

\begin{figure}[t]
    \centering
    \begin{subfigure}{0.22\textwidth}
        \centering
        \includegraphics[width=\textwidth]{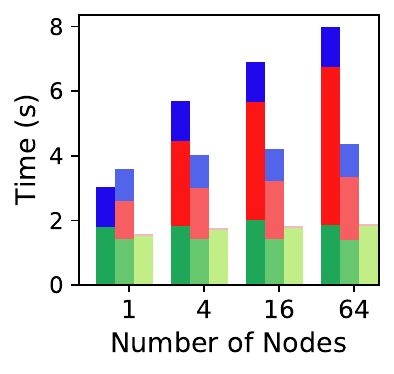} 
        \caption{Filter}
    \end{subfigure}
    \begin{subfigure}{0.22\textwidth}
        \centering
        \includegraphics[width=\textwidth]{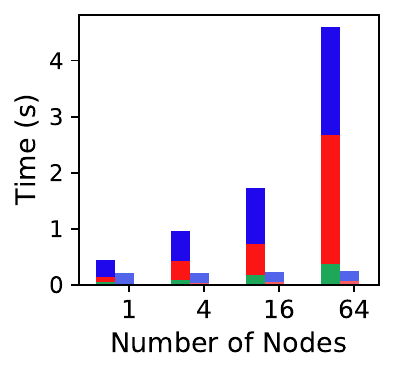}
        \caption{QR}
    \end{subfigure}
    ~\\
    \begin{subfigure}{0.22\textwidth}
        \centering
        \includegraphics[width=\textwidth]{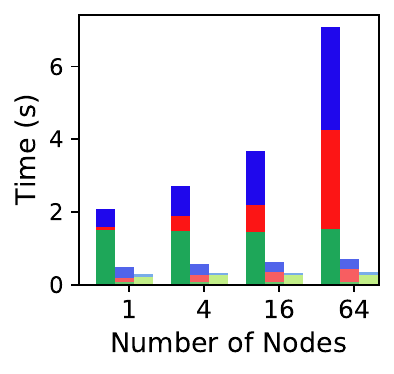}
        \caption{RR} 
    \end{subfigure} 
    \begin{subfigure}{0.22\textwidth}
        \centering
        \includegraphics[width=\textwidth]{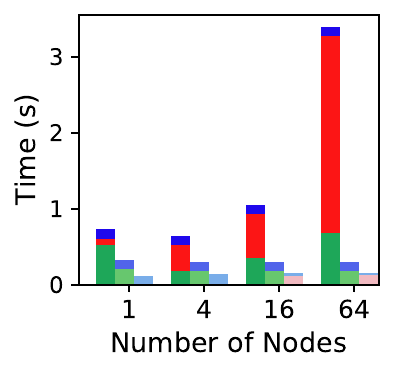} 
        \caption{Residual}
    \end{subfigure}
    
    \caption{Computation (marked in green), communication (red) and data movement (blue) costs within \textsc{Filter}, \textsc{QR}, \textsc{Rayleigh-Ritz} (RR) and \textsc{Residual} of ChASE(LMS) (brightest color shades), ChASE(STD) and ChASE(NCCL) (lighter color shades).}\label{fig:Computation vs Communication}
\end{figure}

The results are shown Fig. \ref{fig:Computation vs Communication}, where stacked bar plots are used to show the portion of computation (marked in green), communication (marked in red) and data movement (marked in blue). We use different color shades to distinguish versions of ChASE. ChASE(LMS) is marked with the brightest colors, ChASE(NCCL) is marked with the lightest color, and ChASE(STD) is in middle. 

For all the kernels, ChASE(STD) has already obtained significant reduction of the communication overhead compared with ChASE(LMS) except for the {\sf Filter} using only $1$ node; in this case ChASE(LMS) takes advantage of a configuration with 4 GPUs per MPI rank. Furthermore, the data movement has been fully removed in ChASE(NCCL), and the overhead of collective communications provided by NCCL is negligible compared with the overhead of communication in ChASE(STD) based on MPI.

The runtime of computation and communication in ChASE(LMS) increases substantially for larger number of computing nodes and matrix sizes, while the new ChASE can maintain a good weak scaling performance, especially for ChASE(NCCL). In summary, ChASE(STD) attains respectively speedups of $1.6\times$, $22\times$, $10\times$, $8\times$ over ChASE(LMS) with $64$ compute nodes on JUWELS-Booster for the {\sc Filter}, {\sc QR}, \textsc{Rayleigh-Ritz} and \textsc{Residuals}. Meanwhile, ChASE(NCCL) attains respectively speedups of $3.8\times$, $1149\times$, $23\times$, $33\times$ over ChASE(LMS), and speedups of $2.3\times$, $51\times$, $2.2\times$, $4\times$ over ChASE(STD). The speedup of QR in ChASE(NCCL) is extremely large as only CholeskyQR2 is employed which does not have any data movement at all. This is also reflected by the practically invisible bars in fig. \ref{fig:Computation vs Communication}b. 

\subsection{Scalability}

In this last part, we describe the behavior of ChASE(NCCL) in the weak and strong scaling regimes and compared it with both ChASE(LMS) and ChASE(STD).

\begin{figure}[t]
    \centering
    \begin{subfigure}{0.46\linewidth}
        \centering
        \includegraphics[width=.99\linewidth]{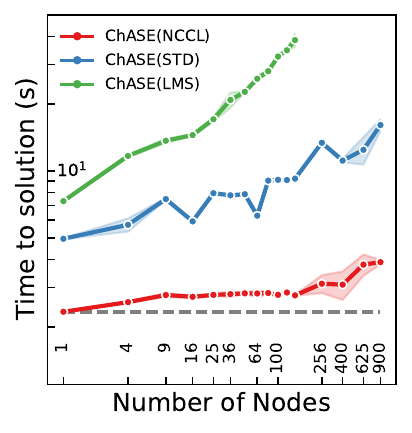} 
        \caption{Weak scaling \label{fig:weak_scaling}}
    \end{subfigure}    
    \hspace{10.pt}
    \begin{subfigure}{0.46\linewidth}
        \centering
        \includegraphics[width=.99\linewidth]{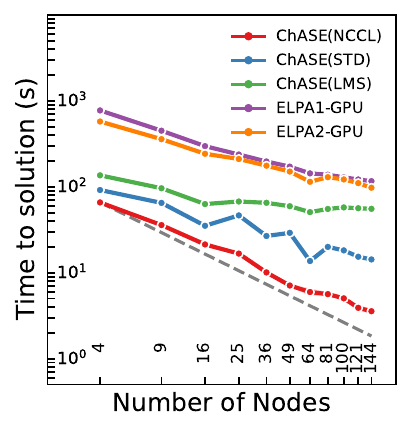} 
        \caption{Strong scaling \label{fig:strong_scaling}}
    \end{subfigure}
    
    \caption{Strong and weak scaling tests with  {\sf Uniform} matrices. Each data point is the result of 8 repetitions.}
\end{figure}

\subsubsection{Weak scaling}\ Weak scaling experiments are particularly important to domain scientists, because they represent the potential of a library to execute computations of systems of increasingly larger size. For the weak scaling tests, we employed up to 900 
of the JUWELS-Booster, and a total of $3,600$ NVIDIA A100 GPUs. The compute nodes count is chosen to be a square integer $1,4,9, \cdots,144,256,400, 625, 900$ to ensure square 2D MPI grids.

For all weak scaling tests, we used artificial matrices of type {\sf Uniform}, with a size increment of $30$k ($30$k, $60$k, $90$k, $\cdots$). 
The maximal matrix size tested is $900$k. {\sf nev} and {\sf nex} were fixed to $2,250$ and $750$. Only a single iteration of ChASE has been executed for all the experiments of weak scaling to ensure a fixed workload per task. This is because being an iterative method, ChASE might require different number of iterative steps for matrices with increasing size. 

Fig. \ref{fig:weak_scaling} shows that the weak-scaling behaviour of ChASE(NCCL) is close to optimal: when the matrix size increases from $30$k to $900$k, 
the time-to-solution for a single iterative step increases by a factor of $1.8$ from $2.3$s to $3.9s$. With the novel designed parallelization scheme and algorithm, ChASE achieves much better weak-scaling performance even without NCCL and CUDA-awareness; 
the time-to-solution of ChASE(STD) increases only by a factor of $3.1$ from $5.1$s to $16$s. 
There are some special points on the curve of ChASE(STD) (i.e., the cases with compute nodes count equal to $4$, $16$, $64$ and $256$) where the time-to-solution drops down. 
The improvement in performance on these points is caused by the MPI\_Allreduce collective communication, within the row or column communication, which is implemented based on a binary tree scheme. When the numbers of MPI ranks in the row or column communicator is a power of $2$, it offers an advantage to these configurations over the others.

The improvement over ChASE(LMS) is quite evident. The weak-scaling experiments of ChASE(LMS) could only test up to $144$ nodes, because of the large memory footprint \cite{wu2022chase}. ChASE(NCCL) and ChASE(STD) on $144$ compute nodes reaches a $14.1\times$ and $4.6\times$ speedup over ChASE(LMS), respectively. 

\subsubsection{Strong scaling}\ Fig. \ref{fig:strong_scaling} illustrates the results of the strong scaling experiments using the In$_2$O$_3$ $115$k eigenproblem, listed in Table \ref{tab:real_world matrix}. The number of eigenpairs sought after is set at $1,200$, representing $\sim 1\%$ of the full spectrum. The size of the external searching space {\sf nex} is fixed as $400$. As a reference, the strong-scaling performance of ChASE is compared with ELPA \cite{ELPA2014EigenvalueELPA}, the state-of-the-art eigensolver for solving dense Hermitian eigenproblems on distributed-memory heterogeneous systems. The version of ELPA used is 2022.11.001.rc1, compiled with the same software stack for ChASE. In ELPA, the block size of the block-cyclic distribution is fixed at $16$. Data are obtained with $8$ repetitions. The compute nodes count are selected to be square of integers $4, 9, \cdots, 121, 144$.

ChASE(STD) featuring the novel parallel scheme attains already much better strong-scaling performance than ChASE(LMS). The time-to-solution of the former drops from $~\sim 92$s on 4 nodes to $\sim14$s on 144 nodes achieving $6.6\times$ speedup. On the contrary, the time-to-solution of ChASE(LMS) only decreases from $~\sim135$s on 4 nodes to $~\sim 55$s on 144 nodes gaining only a speedup of $2.5\times$  


The strong-scaling performance achieved by ChASE(NCCL) is close to ideal. Comparing the execution on $4$ and $144$ compute nodes, ChASE(NCCL) achieves $18.6\times$ speedup, with the time-to-solution dropping from $~\sim 65$s to $\sim3.5$s. When $4$ nodes are utilized, ChASE(STD) and ChASE(NCCL) achieve $1.5\times$ and $2.09\times$ speedup over ChASE(LMS), respectively. These speedup are enlarged to $3.9\times$ and $15.7\times$ when $144$ nodes are utilized.
 
Conversely, ELPA1-GPU and ELPA2-GPU display only $6.7\times$ and $5.9\times$ speedup. When comparing ChASE with ELPA, ChASE(NCCL) experiences an increase in the speedup over ELPA2-GPU to reach
a $28\times$ speedup, as the node count increases. On $144$ compute nodes---$576$ NVIDIA A100 GPUs on JUWELS-Booster---ELPA2-GPU computes the $1,200$ exterior eigenpairs of the $115$k Hermitian dense eigenproblems in $~\sim98$s, while ChASE(NCCL) takes $~\sim3.5$s. We want to emphasize that the performance gain of ChASE over ELPA is obtained when only a relatively small portion of exterior eigenpairs are desired, which is the target usage of the ChASE. Despite being justified, such choice may put ELPA at a disadvantage. 
\section{Conclusion}\label{sec:conclusion}

In this paper, we present a number of major improvements carried out on the ChASE library targeting distributed GPU systems for solving large-scale symmetric and Hermitian eigenproblems.  ChASE targets dense eigenproblems when a relatively small fraction ($\leq10$\%) of extremal eigenpairs is sought after. The improvement includes i) a redesign of the buffer structure, ii) a novel parallelization scheme, iii) a mechanism to switch between different variants of communication avoiding QR based on accurate estimates of the condition number of the matrix of vectors outputted by the Chebyshev filter, and iv) a substitution of the MPI library by the NCCL library for collective communications. The accuracy of the estimation of the condition number and the replacement of Householder QR by CholeskyQR has been verified and validated by numerical tests with a series of eigenproblems extracted from  Condensed Matter applications. The parallel performance of the new ChASE v1.4 version has been benchmarked on the supercomputer JUWELS-Booster and is able to attain excellent strong and weak scaling performance. The resulting library can tackle dense problems with size up to $n=\mathcal{O}(10^6)$, and scale up to the full $900$ nodes of JUWELS-Booster comprising $3,600$ NVIDIA GPUs in total. In the future, we plan to port ChASE to AMD GPUs using the RCCL library.


\begin{acks}
The authors gratefully acknowledge the computing time granted by Jülich Supercomputing Centre (JSC)
on JUWELS-Booster. We thank the members of the department of 
High-Performance Computing Systems at JSC for support and coordination of Full-Node Scale benchmarks on JUWELS-Booster.

\end{acks}

\bibliographystyle{ACM-Reference-Format}
\bibliography{reference}


\end{document}